\documentclass[sigconf,nonacm, balance=false]{acmart}

\usepackage{algorithm}
\usepackage{algpseudocode}
\usepackage{array}
\usepackage{ifthen}
\usepackage{makecell}

\setcopyright{none}
\renewcommand\footnotetextcopyrightpermission[1]{}

\begin{document}
\title[BlindFL: Segmented FL with FHE]{BlindFL: Segmented Federated Learning with Fully Homomorphic Encryption}


\author{Evan Gronberg$^*$}
\orcid{0009-0008-6229-2797}
\affiliation{%
  \institution{Leidos Inc.}
  \city{Riverside}
  \state{CA}
  \country{USA}}
\email{evan.m.gronberg@leidos.com}

\author{Liv d'Aliberti$^*$}
\orcid{0000-0002-3139-5960}
\affiliation{%
  \institution{Leidos Inc.}
  \city{Arlington}
  \state{VA}
  \country{USA}}
\email{olivia.daliberti@leidos.com}

\author{Magnus Saebo$^*$}
\orcid{0009-0007-3264-7211}
\affiliation{%
  \institution{Leidos Inc.}
  \city{Arlington}
  \state{VA}
  \country{USA}}
\email{magnus.m.saebo@leidos.com}

\author{Aurora Hook}
\orcid{0009-0002-3892-405X}
\affiliation{
  \institution{Leidos Inc.}
  \city{Nashville}
  \state{TN}
  \country{USA}
}
\email{auri.hook@leidos.com}

\renewcommand{\shortauthors}{Gronberg, d'Aliberti, Saebo, Hook}

\begin{abstract}

Federated learning (FL) is a popular privacy-preserving edge-to-cloud technique used for training and deploying artificial intelligence (AI) models on edge devices. FL aims to secure local client data while also collaboratively training a global model. Under standard FL, clients within the federation send model updates, derived from local data, to a central server for aggregation into a global model. However, extensive research has demonstrated that private data can be reliably reconstructed from these model updates using gradient inversion attacks (GIAs). To protect client data from server-side GIAs, previous FL schemes have employed fully homomorphic encryption (FHE) to secure model updates while still enabling popular aggregation methods. However, current FHE-based FL schemes either incur substantial computational overhead or trade security and/or model accuracy for efficiency. We introduce BlindFL, a framework for global model aggregation in which clients encrypt and send a subset of their local model update. With choice over the subset size, BlindFL offers flexible efficiency gains while preserving full encryption of aggregated updates. Moreover, we demonstrate that implementing BlindFL can substantially lower space and time transmission costs per client, compared with plain FL with FHE, while maintaining global model accuracy. BlindFL also offers additional depth of security. While current single-key, FHE-based FL schemes explicitly defend against server-side adversaries, they do not address the realistic threat of malicious clients within the federation. By contrast, we theoretically and experimentally demonstrate that BlindFL significantly impedes client-side model poisoning attacks, a first for single-key, FHE-based FL schemes.

\end{abstract}


\maketitle

\def\thefootnote{*}\footnotetext{These authors contributed equally to this work}
\def\thefootnote{\arabic{footnote}}
\section{Introduction}

AI models allow computers to process, analyze, and respond to data by detecting patterns and making predictions in a way that mirrors human responses. To develop predictive capability, AI model construction conventionally relies upon the collection of data in a single location. This presents a privacy concern, since the centralization of data could result in the potential misuse of sensitive information such as personally identifiable information (PII).

Federated learning (FL) has been popularized as a way to collaboratively train a shared AI model while keeping training data at the edge. Instead of gathering all data in a centralized location, FL typically relies on edge nodes, or ``clients'', to train models on local data at the edge and then pass their locally trained models to a central server for aggregation into a global model. Then, the newly aggregated global model is sent back to the clients for further fine-tuning on their potentially sensitive data. 
This way user data do not leave the edge device, and only models trained on that data.
Today, AI models are increasingly trained and deployed on edge devices using FL architectures for services like text completion \cite{textFL}, self-driving cars \cite{carFL}, healthcare services \cite{healthFL}, and other domains \cite{generalFL}.

%

Although, the standard FL procedure alone does not guarantee edge data privacy \cite{PPFL}. 
Adversarial attacks--such as gradient inversion attacks (GIAs), membership inference attacks, and property inference attacks--have been shown to compromise the security of client data in FL systems \cite{shi2024dealing}. 
Specifically, GIAs can recover data used to train a model only from its gradients, allowing the aggregation server to reconstruct client data from received updates \cite{gradient-pruning}. Moreover, GIAs have been shown to reconstruct complex image and text data from deep neural networks, demonstrating a strong need for further FL security considerations \cite{gradient-pruning, wei2023client, shi2024dealing, byzantine_grad, gradientSurvey, hl-constraints, evaluating-gradient-attacks}.


 
To address GIAs and other server-side privacy risks, several secure aggregation procedures have been developed as privacy-preserving federated learning (PPFL) protocols \cite{secureprivacy, comprehensive-review-ppfl, shi2024dealing}. 
However, prior protocols have fallen short by either sacrificing model accuracy, system speed, or memory for privacy, or by limiting privacy guarantees to minimize costs. Approaches that use differential privacy, such as in the work of Cheng et al. \cite{dp-fed}, sacrifice model accuracy to enhance edge user privacy. Similarly, sanitization-based techniques reduce model accuracy by excluding sensitive features and still remain vulnerable to re-linkage \cite{relinkage}.
Prior work has shown that leveraging fully homomorphic encryption (FHE) for secure aggregation can come with significant associated speed and memory costs \cite{xie2024efficiency}. 
The work of Jin et al. on FHE-based FL reduces the speed and memory costs by selectively encrypting sent model updates, but in doing so sacrifice provably secure aggregation \cite{fhe-fed}.
In addition, recent work has shown how malicious clients within a federation can recover edge data with model poisoning techniques based around GIAs \cite{wei2023client}. Malicious client attacks circumvent the security of single-key FHE aggregation as they work off of the decrypted global model. The significant threat posed by malicious clients has been largely overlooked, and has not been considered by other single-key, FHE-based FL schemes, to the best of our knowledge \cite{shi2024dealing}.

In this paper, we propose a new federated learning architecture, BlindFL, which marries FHE-based FL with a model segmentation technique, called client model segmentation (CMS), for efficient and secure FL.
With CMS, only subsets of the edge models are encrypted and shared with the server, thus reducing the total amount of data that is encrypted/decrypted. For clarity, we ensure that the server only aggregates encrypted model segments, and that the global model is only decrypted by clients.
Choice over how much of the client models is encrypted and sent allows for flexibility between optimizing for system performance and for model performance.
Though, we show that model performance is preserved even when client models only send small subsets of their model for aggregation.
Moreover, we show experimentally and theoretically that CMS provides substantial protection from current malicious client attacks with proper choice of CMS parameters. 
Our work demonstrates that BlindFL can be significantly more computationally efficient than plain FHE-based FL while both maintaining model performance for varied client counts and providing additional client data security.


\textbf{Contributions.} We propose BlindFL, a scalable PPFL architecture that makes use of both FHE and an efficient aggregation method, client model segmentation (CMS). We then describe the standard FL protocol, detail client-side GIA risks, and provide a threat model. Next, we provide the algorithms for CMS, FHE key management, and the system architecture. We carry out extensive experimentation using the MNIST \cite{mnist} and CIFAR-10 \cite{cifar10} datasets, examining the relationship between metrics such as accuracy, timing, and data transfer with respect to rounds of BlindFL. Additionally, we demonstrate, both theoretically and experimentally, that CMS defends against malicious client-to-client, intra-federation adversaries.

\section{Preliminaries}

We start by discussing the basic FL model, and then provide an overview of the threat model and potential privacy attacks. 

\subsection{Federated Learning}

\begin{figure}[h]
  \centering
  \includegraphics[width=0.8\linewidth]{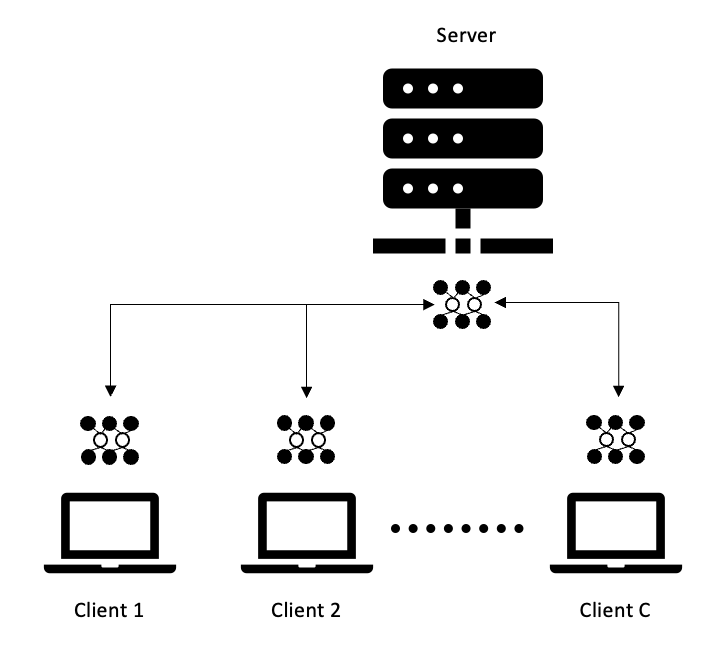}
  \caption{Basic diagram of standard, centralized FL.}
  \label{fig:fl_diagram}
\end{figure}

FL allows multiple data owners to train models such that each data owner's data remains siloed. Centralized FL uses a one-to-many, server-client architecture, with $C > 1$ clients (i.e., data owners) and 1 central server orchestrating the clients (i.e., the federation). Each client $c_i$ contains a local model with parameter matrices $w_i$ that can be trained based upon observations $D_i$. In Figure \ref{fig:fl_diagram}, we show a basic diagram of centralized FL, whereby the server creates an untrained model and sends the model's parameters, $W$, to all clients within the federation. Then, $c_i$ initializes its model with $W$ and trains over $D_i$ to produce local model parameters $w_i$. Each $c_i$ then sends the trained $w_i$, along with the size of its training set $t_i = |D_i|$, to the server.

Our approach is built upon the FedAvg scheme, a weight averaging approach to FL \cite{FL}. To update the global model under FedAvg, the server carries out the following aggregation procedure:

$$
W_{\text{new}} = \frac{\sum_{i=1}^{C} w_i t_i}{\sum_{i=1}^{C} t_i}
$$

$W_{\text{new}}$ is then sent to all clients, and this concludes a round of FL. Successive rounds may be conducted to produce improved models. Variations from this process can be found in Moshawrab, et al. \cite{fl_algos}. The primary privacy-preserving feature of FL is that each client gets the benefit of a model trained on every client's data without having to reveal local data to any party.

\subsection{Gradient Inversion Attack Threat}
There is a flaw in the privacy-preserving protections offered using standard FL. The privacy of FL is provided by the protocol's ability to prevent client data from leaving the edge device. However, sending full, plaintext models to the server poses a serious risk: the server, or a malicious client, could reverse engineer the model gradients to potentially reveal sensitive characteristics of the data, such as PII, used to train client models. This risk is well-researched \cite{fl-attacks, shi2024dealing, PPFL, comprehensive-review-ppfl}.

GIAs aim to reconstruct training data from the gradients of a model \cite{hl-constraints, inversion-attacks-graph-nn, evaluating-gradient-attacks}. More recent attacks have been show to accurately reconstruct multiple data points with minimal knowledge of the client training dataset \cite{evaluating-gradient-attacks, inverting-gradient-attacks-easy}. This type of attack poses a serious threat to data privacy, rendering training data vulnerable to exposure based only upon model gradients. Attempts to mitigate the threats posed by this type of attack have had varying results. Attempted protections have included gradient pruning \cite{gradient-pruning}, the \textit{mixup} principle \cite{mixup}, and InstaHide \cite{instahide}. Each comes with inherent risks and benefits, and each offers varying degrees of protection as evaluated by Huang, et al. \cite{evaluating-gradient-attacks}.

\subsection{Threat model}

\begin{figure}[h]
  \centering
  \includegraphics[width=\linewidth]{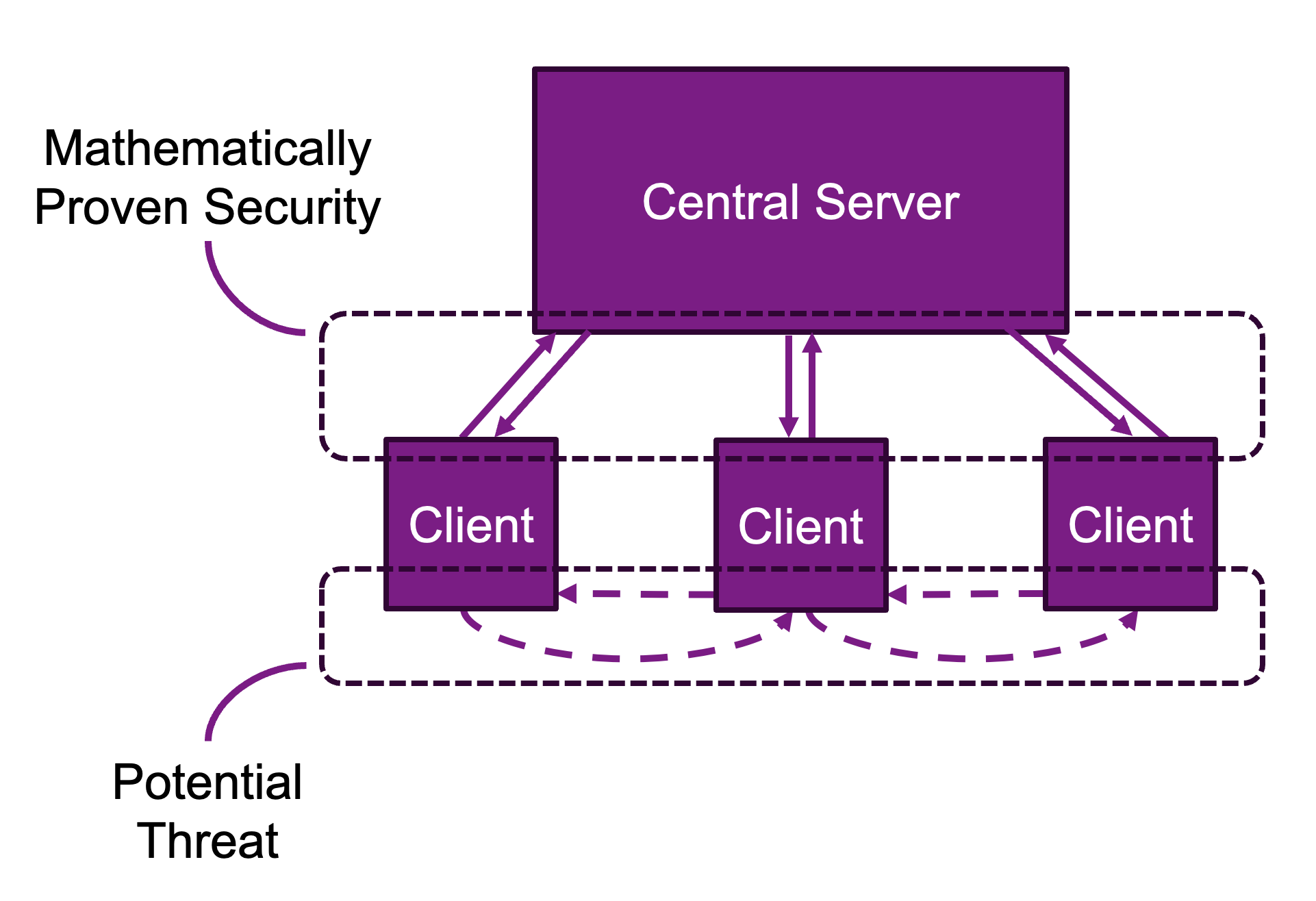}
  \caption{Diagram displaying the security relationship between client and central server within FHE-based FL. FHE ensures secure aggregation by the central server, but does not defend against malicious clients.}
  \label{fig:threat_diag}
\end{figure}

We define a malicious adversary $\mathcal{A}$ that has white-box access to either the server or a subset of clients within a BlindFL federation.
Unlike ``honest-but-curious'' adversaries, which do not interfere with model updates and adhere to the FL protocol defined in Section 2.1, $\mathcal{A}$ is ``malicious'' and therefore can alter model updates and stray from the FL protocol to obtain privileged information that is less obtainable during the normal FL process. We explain in Section 5 that FHE encryption of model updates provably secures local data from server-side adversaries. 
In the same section, we highlight the threat of private data leaking between clients in the federation, as illustrated in Figure \ref{fig:threat_diag}. 
As such, we primarily focus on the scenario where $\mathcal{A}$ compromises $c$ out of $C$ clients in the federation for $k$ rounds instead of $\mathcal{A}$ compromising the central server.

Given that $\mathcal{A}$ compromises $c$ clients, $\mathcal{A}$ may, for any given round $i$, decide to send maliciously crafted updates for aggregation for any of the $c$ clients with the intention of poisoning the global model. Specifically, $\mathcal{A}$ aims to poison the global model to recover data samples from the local training datasets of uncompromised clients, as outlined in prior literature for malicious clients \cite{wei2023client}. We assume that the only information $\mathcal{A}$ has of the underlying dataset for a FL task is the information present in the seized $c$ datasets, including estimates as to the mean and standard deviation of the underlying data distribution as these values are necessary for our chosen GIA \cite{inverting-gradient-attacks-easy}. To test the limits of our novel federation scheme, we assume capabilities of $\mathcal{A}$ which surpass the state-of-the-art for client-based, malicious adversaries. For our work, we assume that $\mathcal{A}$ can poison the global model to the extent that $\mathcal{A}$ can exactly recover the update gradients corresponding to data samples of interest. $\mathcal{A}$ then reconstructs the private data samples used to generate the isolated update gradient with a GIA.

In addition to proving security between client data and the central server, we show that even for extreme capabilities of $\mathcal{A}$, BlindFL provides flexible security between client data and malicious clients.

\section{Related Works}

FL was first proposed by McMahan, et al. as a way to privately and efficiently train models deployed on mobile devices \cite{FL}. Several frameworks have since been developed to implement FL in a variety of ways that further enhance privacy. The seminal work describing PPFL implementations in PyTorch was written by Ryffel, et al. \cite{PPML}. Our own PPFL, BlindFL, is built on top of Flower, an accessible and extensible FL framework for privacy-preserving research \cite{flower}.

As FL implementations grew in popularity, so did the number of attacks proposed upon FL edge-to-cloud systems. Inferencing attacks resulting in information leakages was first proposed by Hitaj, et al. \cite{Hitaj}. Further notable works include GIAs mentioned in Wang, et al. \cite{hl-constraints} and  He, et al. \cite{inv-attacks-he}. A survey of GIAs is covered under Huang, et al. \cite{evaluating-gradient-attacks}. A survey of malicious adversarial attacks and defenses is described in Shi, et al. \cite{shi2024dealing}.

Various countermeasures to defend against attacks to FL systems have also emerged. In 2019, Zhu, et al. proposed gradient pruning as a method to obscure model gradients without any changes in training \cite{gradient-pruning}. In 2020, Huang, et al. proposed InstaHide, which encrypts all training images with one-time secret keys \cite{instahide}. Though, a 2021 review of PPFL techniques by Yin, et al. note that several security approaches result in a decrease in accuracy \cite{comprehensive-review-ppfl}. Hu, et al. worked on decentralized FL with a segmented gossip approach, and their ``segmented aggregation'' method served as inspiration for our CMS method \cite{gossip-approach}. To our knowledge, BlindFL is the first to use CMS without decentralization.

In 2018, Phong, et al. used homomorphic encryption (HE) to protect deep learning model gradients \cite{fhe-deep-learning}. In 2023, Rahulamathavan, et al. worked on the FheFL, applying FHE directly to FL model gradients \cite{fhefl-data-defense}. 
In 2023, Jin, et al. proposed FedML-HE, an approach that selectively encrypts only the most privacy-sensitive parameters within a client model, mitigating the costs of FHE. 
Unlike the FedML-HE approach, we maintain fully encrypted aggregation while also reducing computational resource use.
In 2022, Sébert, et al. explored combining HE with differential privacy for protecting FL training data \cite{he-data-defense}. With our approach combining FHE and CMS, we are able to converge without compromising accuracy.

Research is also developing methods for detecting information leakage. A sensitivity metric has been developed to quantify the information leaked by a model gradient by Mo, et al. in 2021 \cite{mo2020layer}. Elkordy, et al. produced a paper in 2022 outlining the theoretical bounds for information leakage and its relationship with the number of clients participating \cite{fl-secure-aggregation}.

The literature on federation adversaries has identified powerful malicious client attacks which can isolate and reconstruct target data samples out of global model updates aggregated over 1000 clients \cite{wei2023client}. Defenses for such malicious client adversaries remain sparse, and the techniques provided within the FHE-based FL literature require multi-key encryption, which increases the computational overhead for FL compared to single-key \cite{park2022HEppfl, aziz2023explore, fhefl-data-defense}. While single-key schemes have employed differential privacy (DP) for added security, DP has been shown to be ineffective against malicious clients \cite{fhe-fed, wei2023client}. To our knowledge, no other single-key, FHE-based FL scheme addresses the threat of malicious client adversaries. Meanwhile, the CMS aggregation method limits the possible information leaked about any one local dataset in the aggregated global model. Moreover, we experimentally show that malicious clients leveraging state-of-the-art GIAs are markedly less effective with moderate use of CMS. In this way, BlindFL offers security unparalleled by related methods.



\section{Methods}

BlindFL addresses the threat of GIAs uncovering sensitive data by combining two methods:

\begin{itemize}
    \item \textbf{Client Model Segmentation (CMS)}: Instead of each client sending its full model, the server only requests a pre-determined subset of parameter matrices from each client model.
    \item \textbf{Fully Homomorphic Encryption (FHE)}: Client model segments are homomorphically encrypted, using CKKS \cite{CKKS}, and then are sent to the server. FHE enables computations to be run on encrypted data, so the server is still able to aggregate received segments.
\end{itemize}

Below we describe our system setup, including algorithms and protocols for implementing BlindFL.

\subsection{Client Model Segmentation}

Suppose that there is 1 server and $C \ge 2$ clients, each of which has 1 model. Suppose that each model is a deep neural network (DNN) with $M > 0$ parameter matrices. The server selects $c \le C$ clients with $c \ge 2$. Then, the server generates $c$ random binary sequences $\{b_j\}$ of length M, where $j = 1, 2, .., M$, which we represent as request matrix $R$ of size $c \times M$. The server generates $R$ such that the following property holds true:

$$
\sum_{i=1}^{c} R_{i}^{(j)} \ge p, \forall j = 1, 2, ..., M
$$

where $p$ is the number of client parameter matrices that must be gathered for each server-side global parameter matrix and $j$ iterates over the parameter matrices of client $c_i$'s model. The value $p \ge 1$ is configured before BlindFL begins. Further, since each element of $R$ is either 0 or 1, it is clear that $p \le c$.

To generate $R$ such that the above property holds true, the server calculates the number of matrices each client should send back, defined by $N$ below:

$$
N = \lceil \frac{M \cdot p}{c} \rceil
$$

For the first selected client, the server generates a sequence, $R_1$, of $M$ booleans where $N$ random booleans are set to 1. Then, for each following $R_i$, the server initializes a sequence of $M$ booleans all set to 0. It next calculates the sum of all previously generated $R_i$ sequences in the form of an $M$-length sequence where each value in the sequence denotes how many client parameter matrices will currently be requested for that given global model parameter matrix. Then, $N$ times, the minimum value in that summation is found, and a random index that has that minimum value is selected. A 1 is inserted at that index into $R_i$. As 1s are inserted, the summation is updated. The generation of these request matrices is formalized in Algorithm \ref{algo:1}.

\begin{algorithm}
\caption{Request Matrix Generation}
\begin{algorithmic}[1]
\Procedure{Generate-Request-Matrix}{$M, c, p$}
    \State {$N = \lceil \frac{M \cdot p}{c} \rceil$}
    \State {$R \gets $ array of $c$ empty arrays of size $M$}
    \State {$R_1 \gets$ randomly update with $N$ $1$'s}
    \For {$i \gets 2$ to $c$}
        \State{min-ind $= \{\}$}
        \For{$j \gets 1$ to $N$}
            \State{$\textbf{sum}_i \gets$ array of $M$ 0's}
            \For{$k \gets 1$ to $i$}
                \State{$\textbf{sum}_i \gets \textbf{sum}_i$ + $R_k$}
            \EndFor
            \State{min-ind $\gets argmin_i(\textbf{sum}_i)$}
            \State{$R_{i,[min-ind]} \gets 1$}
        \EndFor         
    \EndFor            
    \State \Return $R$
\EndProcedure
\end{algorithmic}
\label{algo:1}
\end{algorithm}

The server then sends to each client the corresponding matrix row. For each $i = 1, 2, ..., c$, where $R_{i,j} = 1$ for any $j = 1, 2, .., M$, client $i$ will send the server its $j^{\text{th}}$ parameter matrix, $w_{i,j}$, as well the number of examples that were used to train its model, $t_{i}$. The weight response matrix generation algorithm is formalized in Algorithm \ref{algo:2}.

\begin{algorithm}
\caption{Response Matrix Generation}
\begin{algorithmic}[1]
\Procedure{Generate-Response-Matrix}{$R$}
    \State{$\text{cm} \gets $ a $c\times M$ matrix of client model weights}
    \State{$w \gets $ an empty array of shape $R$}
    \State{$t \gets $ an empty array of shape $R$}
    \For {$i \gets 1$ to $c$}
        \For{$j \gets 1$ to $M$}
            \If{$R_{i,j} == 1$}
                \State{$w_{i, j}\gets \text{cm}_{i, j}$}
                \State{$t_{i, j} \gets $ count$($examples$)$ by class}
            \EndIf
        \EndFor
    \EndFor
    \State \Return $w,t$        
\EndProcedure
\end{algorithmic}
\label{algo:2}
\end{algorithm}

Once the server has received all $p$ requested parameter matrices, $w_{i,j}$, and the $p$ requested training example counts, $t_{i}$, it creates the global model using the following formula to calculate each aggregated parameter matrix $W_j$:

$$
W_j = \frac{\sum_{i=1}^p w_{i,j} t_{i,j}}{\sum_{i=1}^p t_{i,j}}, \forall j = 1, 2, ..., M
$$

The server aggregation process for each $W_j$ is formalized in Algorithm \ref{algo:3}. All parameter matrices $W_j$ are then sent from the server to all $C$ clients, thus placing the newly updated global model on each client.

\begin{algorithm}
\caption{Server Aggregation Process}
\begin{algorithmic}[1]
\Procedure{Server-Aggregation}{$w, t$}
    \State {$W \gets $ an empty array of size $M$}
    \For{$j = 1$ to $M$}
        \State{$sum_1 \gets$ zero matrix with shape $w_{\cdot, j}$}
        \State{$sum_2 = 0$}
        \For{$i = 1$ to $p$}
            \State{$sum_1 = sum_1 + w_{i,j}\cdot t_{i,j}$}
            \State{$sum_2 = sum_2 + t_{i,j}$}
        \EndFor
        \State{$W_j=\frac{sum_1}{sum_2}$}
    \EndFor
    \State \Return W
\EndProcedure
\end{algorithmic}
\label{algo:3}
\end{algorithm}

\subsection{Homomorphic Encryption and Key Distribution}

BlindFL leverages asymmetric FHE, which requires two keys: a public key and a private key. The public key contains the information required to perform encryptions and encodings. As a result, the public key can be given to any system performing mathematical operations on the FHE values. However, the public key lacks the information to decrypt any encrypted value. For any party to be able to decrypt the model, the private key is needed.

BlindFL incorporates FHE into the CMS process described above at the following points:

\begin{itemize}
    \item The client homomorphically encrypts the requested parameter matrices before sending them to the server.
    \item The server uses homomorphic operations to perform an encrypted, weighted average of the corresponding client model segments.
    \item The server sends the complete, encrypted, updated model to the clients, and the clients decrypt using the private key.
\end{itemize}

These steps require that FHE keys be properly distributed to maximally preserve privacy. To address this need, we introduce a third node type, a key distributor (KD). The KD creates and shares the public key at the start of each round of training. After all clients have sent their requested parameter matrices, the KD releases the private key to each client in the federation. This ensures that a distinct set of keys is used with every round of training. Otherwise, the prior valid private key would carry over from the previous round of training, increasing the ability of clients to intercept and decrypt other clients' models in transit. The process of creating a key requires four initial parameters:

\begin{itemize}
    \item \textbf{Scheme}: The FHE scheme of choice under use within the framework.
    \item \textbf{n}: The polynomial coefficient modulus, which is directly linked to the multiplicative depth needed for a given computation.
    \item \textbf{Scale}: The constant upscale factor for fixed-point values, which is used by CKKS to keep a certain level of precision during encrypted operation, not required if scheme is not CKKS.
    \item \textbf{Qi-Size}: A list of prime numbers of a predefined size that will be multiplied together to generate the variable coefficient modulus. 
\end{itemize}

\begin{figure}[h]
  \centering
  \includegraphics[width=\linewidth]{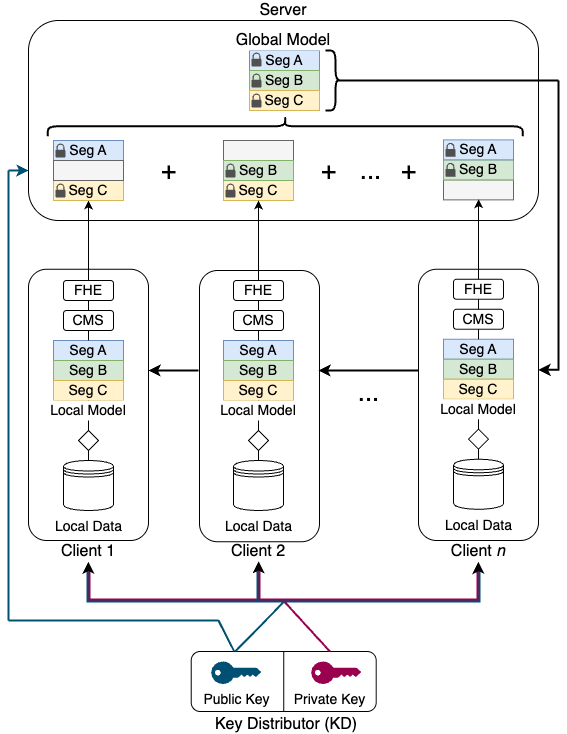}
  \caption{Full diagram of BlindFL.}
  \label{fig:blindfl}
\end{figure}

For more details on the selection of these parameters, refer to Section 6.1. The KD process is incorporated into our system as described in Protocol 1 and shown within Figure \ref{fig:blindfl}.

\begin{algorithm}
\floatname{algorithm}{Protocol 1}
\renewcommand{\thealgorithm}{}
\caption{Key Distributor Round Update Process}
\begin{algorithmic}[1]
\Procedure{Round Updates}{scheme, $n$, scale, qi-size}
    \State {$K_{pub}, K_{priv} = \text{FHE}($scheme, $n$, scale, qi-size$)$}
    \State {$R$ = Generate-Request-Matrix$(M, c, p)$}
    \For {$i \gets 1$ to $c$}
        \State {$c_i$ receives $K_{pub}$ from KD}
        \State {$c_i$ receives $R_{i}$ from server}
        \For{$j \gets 1$ to $M$}
            \State{$\text{Enc}(w_{i,j}) \gets K_{pub}(w_{i,j})$}
        \EndFor
        \State{$c_i$ sends $\text{Enc}(w_{i})$, $t_{i}$, $K_{pub}$ to server}
    \EndFor
    \State{KD sends $K_{priv}$ to $c_i$, $\forall i \in \{1,\dots, c\}$}
    \State{$\text{Enc}(W) \gets $ Server-Aggregation$(\text{Enc}(w), t)$ }
    \State{Server sends $\text{Enc}(W)$ to $c_i$, $\forall i \in \{1,\dots, c\}$}
    \For {$i \gets 1$ to $c$}
        \State{$W = K_{priv}(\text{Enc}(W))$}
    \EndFor
\EndProcedure
\end{algorithmic}
\label{algo:4}
\end{algorithm}

We note that during Protocol 1, the server, rather than the client, notifies the KD of completed server aggregation. This protocol is put in place so that only one party has to be trusted to truthfully attest to the completion of the global model. Once the server performs Algorithm \ref{algo:3}, the server alerts the KD to release the private keys.

\section{Privacy by BlindFL}

Consider our single-key, FHE-based FL system where there is a private-public key pair generated. Past work developing FHE-based FL systems have proven that this single-key setup achieves 0-differential privacy between a client and a central server, i.e. no client information is leaked to the server \cite{fhe-fed, fhe-deep-learning}. This ensures complete protection against server-side inversion attacks regardless of whether an honest-but-curious or malicious adversary compromises the central server as the central server only receives encrypted model updates \cite{zhang2020batchcrypt, wei2023client}.

Yet, there is also the question of the security of the client data represented in the global model. Since all clients receive the same global model after decryption, each client has access to a complete aggregation of every client's model updates. This poses a substantial security threat. New work demonstrates that malicious clients can, from the global model update, isolate the portion of the update corresponding to some target samples \cite{wei2023client}. Once the gradient contribution for some samples is isolated from the global model, the malicious client can leverage the wealth of powerful GIAs to recover the target samples \cite{shi2024dealing}. 

The threat of malicious clients has gone largely unaddressed by similar defensive FL approaches, which oftentimes focus solely on honest-but-curious adversaries \cite{shi2024dealing, wei2023client}. By contrast, our CMS procedure disrupts both the model poisoning and the gradient isolation processes present in state-of-the-art malicious client attacks \cite{wei2023client}. Furthermore, we leverage prior work on a sensitivity metric quantifying private information leakage in model gradients to show a linear relationship between CMS and GIA performance degradation.

We now define the malicious client threat and highlight how our CMS procedure thwarts it. Let $\mathcal{A}$ be an adversary as defined in Section 2.3. For round $i$, $\mathcal{A}$ receives global model $M_i$ and constructs a malicious update designed to isolate the gradient contributions for data belonging to some target class $a$ of the main task. $\mathcal{A}$ then sends this update to the central server. For all uncompromised clients $\{c_s\}$ with local data $\{d_j\}$ with ground truth label $a$, after $\{c_s\}$ receive the poisoned model $M_{i+1}$, $\{c_s\}$ finetune their local models with $\{d_j\}$, and in doing so compromise the security of $\{d_j\}$ to $\mathcal{A}$.

When $\mathcal{A}$ receives the next global model $M_{i+2}$, after taking the difference of $M_{i+1}$ and $M_{i+2}$ to get the update gradients $G$ for round $i+1$, $\mathcal{A}$ can isolate the portion of $G$ corresponding to class $a$, $G_a$. This is done by poisoning the aggregated model to perform substantially worse for this class, such that the gradients for this class dominate all other gradients \cite{wei2023client}. With access to $G_a$, $\mathcal{A}$ can employ a GIA to reconstruct the inputs $\{d_j\}$, thus obtaining data local to $\{c_s\}$. Note that this attack has been shown to work with sizes for $\{d_j\}$ in the hundreds \cite{wei2023client}. 

\begin{figure}[h]
  \centering
  \includegraphics[width=\linewidth]{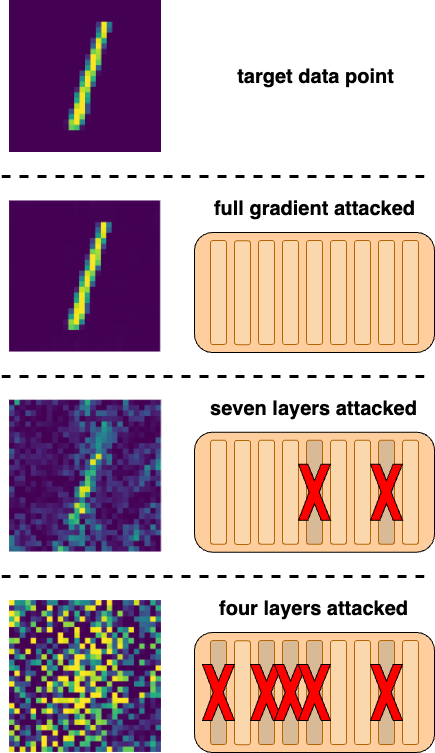}
  \caption{Example of the impact of a GIA on a attacked gradient with a varying number of model layers represented from a ConvNet model trained on MNIST.}   
  \label{fig:inv_diag}
\end{figure}

Now, assume this malicious client attack is happening within a BlindFL federation, where each client sends $n$ layers of their $N$-layer model, with $n<N$. CMS interrupts the construction of the malicious update as it limits $\mathcal{A}$ to use only $n$ layers for the update; therefore, limiting the extent to which $\mathcal{A}$ can alter $M$. Though, let us assume the case where $\mathcal{A}$ can fully circumvent this disruption. Even then, CMS interrupts the success of a GIA by limiting $G_a$ to be a subset of the full-model gradients for data $\{d_j\}$, as $\{c_s\}$ only send $n$ layers of their model update to the central server. We experimentally demonstrate in Section 6.6 that GIA performance is significantly hindered when carried out over gradient subsets. A demonstration of this negative impact on GIA performance can be seen in Figure \ref{fig:inv_diag}. As layer gradients are removed from the attacked gradient, there is a visible degradation in the quality of the GIA reconstruction.

Intuitively, if only a subset of the full gradient contribution for any data point in $\{d_j\}$ is visible to $\mathcal{A}$, then there is a reduction in the total information present about $\{d_j\}$ in $G_a$. We can formalize this intuition with the help of a sensitivity metric designed by Mo, et al. \cite{mo2020layer} which quantifies the amount of leakage of sensitive information in a gradient. This metric is a function of the Jacobian matrix for a gradient, with respect to a loss function, whereby higher "sensitivity" for a parameter indicates more information about the underlying features of a dataset. The leaked information can be exploited by an adversary using a GIA. This correlation between the sensitivity of a gradient and the success of a corresponding GIA has been shown \cite{mo2020layer, fhe-fed}.


The metric is calculated and normalized per-layer, normalizing also by layer size, to allow for comparison between layers. Consider the random variables $\{x_i\}$ which correspond to the sensitivity of layers $1\leq i\leq N$ of a local model $c_s$. If we assume that each $x_i$ is identically distributed with mean $\mu$, then by linearity of expectations, we get that the expectation of the total sum $S$ of $\{x_i\}$ is 

$$\mathbb{E}[S]=\sum_{i=1}^N\mathbb{E}[x_i]=\sum_{i=1}^N\mu=N\mu$$ 

Where, if we now limit the number of layers present in the sum $S'$ to $n < N$, we get the following expectation for a sum of the sensitivity, 

$$\mathbb{E}[S']=\frac{n}{N}\mathbb{E}[S]=n\mu$$ 

Therefore, the total sensitivity of an attacked gradient in expectation is linearly related to the number of layers in the subset gradient. This implies that as $n$ decreases -- as the number of layers sent by a client in the CMS procedure decreases -- the success of gradient inversion attacks also decrease. Though, Figures \ref{fig:attack_1} and \ref{fig:attack_2} indicate a stronger, exponentially decaying relationship between $n$ and GIA performance. This stronger suggested relationship may be due to the fact that the sensitivity metric is per-layer, and thus, does not consider inter-layer connections. These inter-layer connections may be important for GIAs to leverage, and they are the primary way in which CMS disrupts the centralization of private client information within $M$.

\section{Results}

We carried out experiments both to provide evidence of BlindFL's client-to-client security protection and to evaluate protocol system impact. BlindFL performance experiments were run via simulation. The following parameters were set for each run for Sections 6.1-6.5 unless otherwise specified:

\begin{itemize}
    \item \textbf{Client Count}: The static dataset for a given experiment is randomly shuffled, and then evenly partitioned among the clients, i.e. as the number of clients increases, the amount of data provided to each client decreases.
    \item \textbf{Client Parameter Matrix Count}: We vary the number of layers aggregated per global model layer. When this parameter is equal to the number of clients, CMS is effectively inactive.
    \item \textbf{Rounds of FL}: We carry out our experiments over several rounds, see Section 2.1 for round definition. 
    \item \textbf{FHE}: We vary the use of FHE, as well as, the parameters for the FHE encryption, see Section 4.2 and 6.1 for details.
    \item \textbf{CMS}: We vary the use of CMS, and if using CMS, the number of layers each client will share with the global model.
\end{itemize}

The experiments were run on two use cases: (1) classification of the MNIST dataset \cite{mnist} using a LeNet-5 model \cite{lenet} and (2) classification of the CIFAR-10 dataset \cite{cifar10} using a ResNet-20 model \cite{resnet20}. The LeNet-5 model has 10 parameter matrices, defined as all of the weight and bias tensors in the model. See Table \ref{table:1} for a listing of the model parameters and the total bytes consumed by elements in each array. The ResNet-20 model has 128 parameter matrices. If the parameter matrix is requested via Algorithm \ref{algo:1}, then the matrix is encrypted using the public key generated during the round. Each parameter matrix is individually encrypted using the keys sent during Protocol 1.

If a given encrypted parameter matrix $l$ in model $M$ is too large to fit within the protobuf file size limit of 2GB, it is broken into $n$ uniformly sized slices $s_i$ such that $l = [s_1, s_2,..., s_n]$. We then encrypt and share those slices, rather than the full parameter matrix, to be used in Algorithm 3 on the server side.

\begin{table}[h!]
\begin{center}
\begin{tabular}{ |c|c|c| }
 \hline
 Network Neural Layer & LeNet-5 Parameters & Bytes Consumed\\
 \hline\hline
 conv1.weight & 150 & 728\\ 
 conv1.bias & 6 & 152\\  
 conv2.weight & 2,400 & 9,728\\  
 conv2.bias & 16 & 192\\ 
 conv3.weight & 48,000 & 192,128\\  
 conv3.bias & 120 & 608\\   
 fc1.weight & 493,920 & 1,975,808\\ 
 fc1.bias & 84 & 464\\  
 fc2.weight & 840 & 3,488\\  
 fc2.bias & 10 & 168\\ 
 \hline
 Total & 545,546 & 2,183,464\\  
\hline
\end{tabular}
\end{center}
\caption{Diagram of the LeNet-5 model parameter matrix sizes broken down by matrix.}
\label{table:1}
\vspace*{-\baselineskip}
\end{table}
 
All MNIST experiments were run on an AWS EC2 c5.18xlarge instance, which has a vCPU count of 72 and 144 GiB of memory. All CIFAR-10 experiments were run on an AWS EC2 c5.24xlarge, which has a vCPU count of 96 and 192 GiB of memory. The high specifications allowed us to run many clients at once in simulation on a single instance. The accuracy recorded in the figures below is measured as a five-trial average of the global model across each client's test set unless otherwise specified. Flower was used for our FL implementation \cite{flower}, and Pyfhel was used for our FHE implementation \cite{pyfhel}. Pyfhel, relies upon Microsoft SEAL's open-source HE library \cite{sealcrypto}.

\subsection{FHE Context Selection}

Before any other experiments were run, simulations with FHE-enabled and CMS inactive were considered to determine the smallest FHE context that maintains model accuracy while also limiting the time and space overhead of 128-bit security. Our FHE framework, Pyfhel, requires that the following parameters be determined: the scheme, the qi sizes, the $n$ value, and the scale \cite{pyfhel-context}.

We use CKKS \cite{CKKS} as our scheme, as it is able to perform approximate homomorphic computations over floating point numbers. Algorithm \ref{algo:3} requires maintaining accuracy over floating point numbers to accurately calculate weight updates. However, other integer-based FHE schemes could be selected, as long as weights are scaled to become integer values (and then rescaled before inference).

The qi size variable represents the length of each prime used in the creation of the variable coefficient modulus. The coefficient modulus is a substantive integer formed by multiplying primes together. Two primes, the outer primes, are required, even if no multiplicative operation is to be performed \cite{he-params}. By general rule of thumb, we chose the first outer prime to be a 60-bit prime to ensure high precision when decrypting \cite{pyfhel-context}. The second outer prime within the coefficient modulus should be as large as the largest prime within your qi size, so it will also be a prime of length 60 \cite{pyfhel-context}. Together, the outer primes determine the precision of the floating point numbers. The number of inner primes selected is dependent upon the multiplicative depth of the given computation. We are using a leveled FHE scheme, and our computation only requires one multiplicative operation, $w_{i, j} \cdot t_{i, j}$. Therefore, we require two additional primes \cite{he-params}, which we select to be of length $40$ to minimize the space/time impact of FHE on a given computation.


Next, we select our $n$ value, which represents the polynomial coefficient modulus. The value is determined by the polynomial modulus degree. With 128-bit security, given a maximally long qi size of $200$ digits, we can select an $n$ value of 16,384, as it falls within the upper bound of coefficient moduli listed within Table 2 of Pan et al. \cite{he-params}. 

Finally, we select scale. The scaling factor is used to specify the precision during conversion from floating-point numbers to fixed-point numbers. Given our polynomial modulus degree of 16,384, and our desired 128-bit security, we select a scale of $2^{20}$ using Table 4 of Pan et al. \cite{he-params}. In summary, our Pyfhel FHE context, for both MNIST and CIFAR-10 datasets, have the following settings:

\begin{itemize}
    \item Scheme: CKKS
    \item $n: 2^{14}$
    \item Scale: $2^{20}$
    \item Qi Sizes: [60, 40, 40, 60]
\end{itemize}

\subsection{Number of Clients}

\begin{figure}[h]
  \centering
  \includegraphics[width=\linewidth]{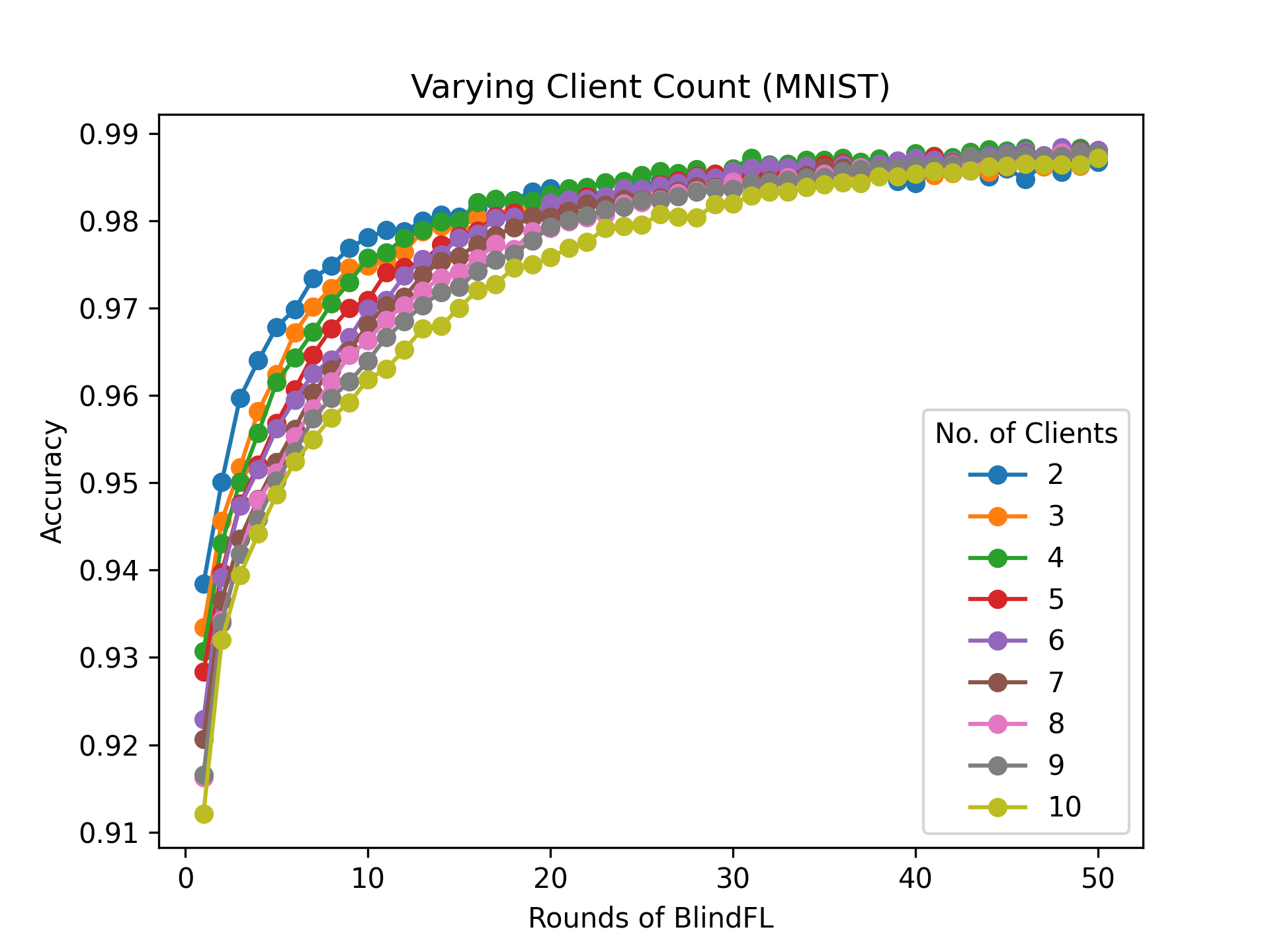}
  \caption{Accuracy averaged over 5 runs, varying client counts and rounds, for a LeNet-5 model trained on the MNIST dataset in a BlindFL federation.}
  \label{fig:mnist_clients}
\end{figure}

\begin{figure}[h]
  \centering
  \includegraphics[width=\linewidth]{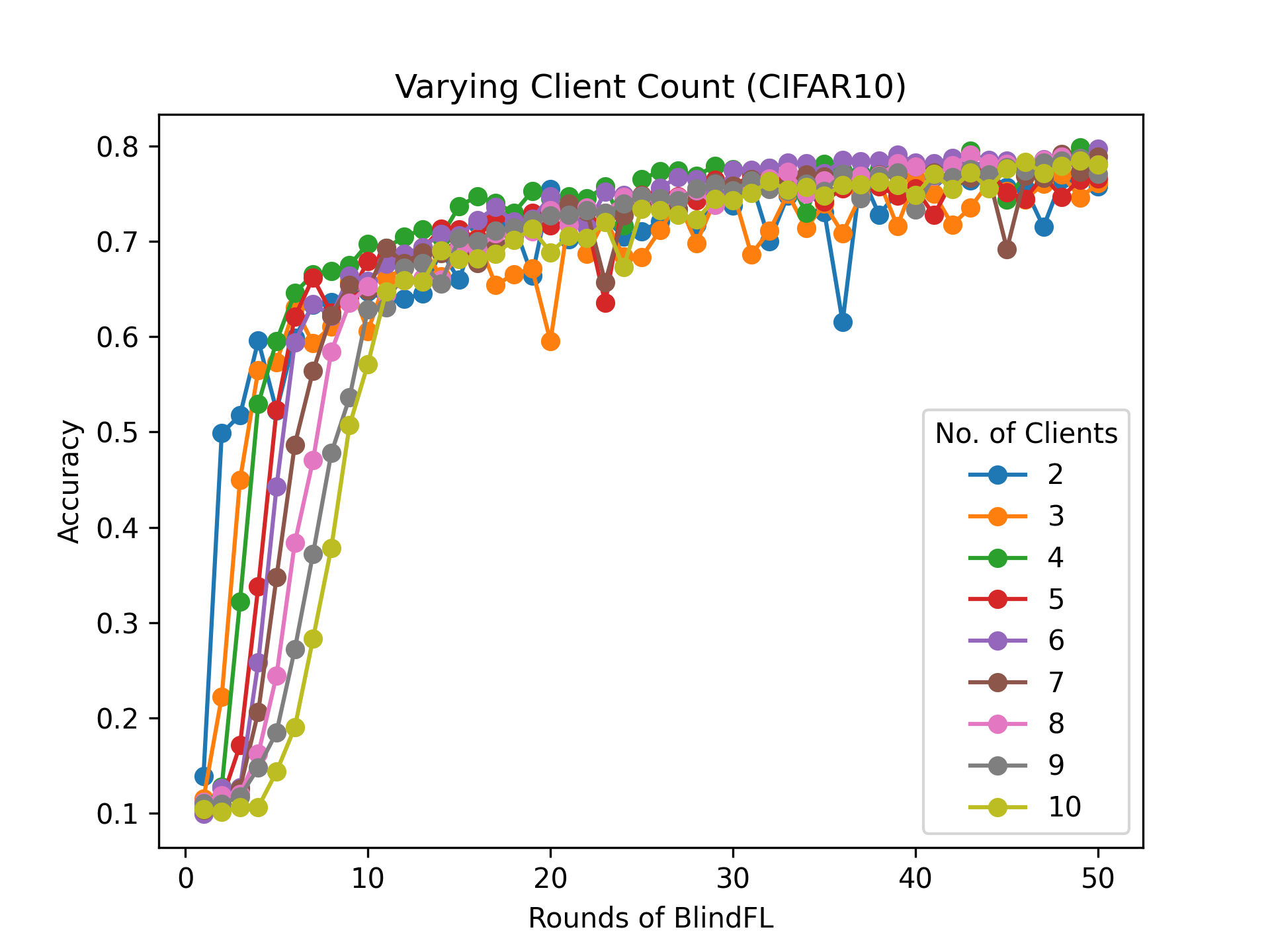}
  \caption{Accuracy averaged over 5 runs, varying client counts and rounds, for a ResNet-20 model trained on the CIFAR-10 dataset in a BlindFL federation.}  
\label{fig:cifar_clients}
\end{figure}

Figures \ref{fig:mnist_clients} and \ref{fig:cifar_clients} show, for both MNIST and CIFAR-10 datasets, how round-over-round performance is affected by changing the number of clients within the federation. Performance is measured as the 5-trial average accuracy of the aggregated global model on each client's test data. This set of experiments was run with both the CMS and FHE components active. For a given global model parameter matrix $j$, we average together $p$ corresponding client model parameter matrices, where $p$ here is equal to the ceiling of half the total number of clients $n$, $p=\lceil\frac{n}{2}\rceil$. So, when we have $10$ clients, for each parameter matrix $j$, $p = 5$, whereas, when we have $2$ clients, $p = 1$. I.e., for each experiment, the global model will always receive half of the number of parameter matrices available.

\begin{table}[h!]

\begin{center}
\begin{tabular}{ |c|c|c| }
 \hline Client Count
  & MNIST Test Accuracy
  & CIFAR Test Accuracy\\
 \hline\hline
 2 Clients & 98.70\% & 76.46\% \\ 
 3 Clients & 98.77\% & 76.35\% \\  
 4 Clients & 98.73\% & 78.27\% \\  
 5 Clients & 98.74\% & 78.54\% \\ 
 6 Clients & 98.75\% & 78.74\% \\  
 7 Clients & 98.76\% & 78.32\% \\    
 8 Clients & 98.81\% & 77.68\% \\ 
 9 Clients & 98.73\% & 77.36\% \\  
 10 Clients & 98.76\% & 77.00\% \\  
\hline
\end{tabular}
\end{center}
\caption{Final recorded accuracies for Figure \ref{fig:mnist_clients} and \ref{fig:cifar_clients}.}
\label{table:2}
\vspace*{-\baselineskip}
\end{table}

The final MNIST and CIFAR-10 test accuracies after 50 rounds of BlindFL are highlighted in Table \ref{table:2}. We note that test accuracy is relatively consistent, even as client counts change. We additionally experiment with holding client count constant to investigate the impact of varying number of parameter matrices contributed per client within BlindFL.

\subsection{Number of Client Model Parameter Matrices per Global Model Parameter Matrix}

\begin{figure}[h]
  \centering
  \includegraphics[width=\linewidth]{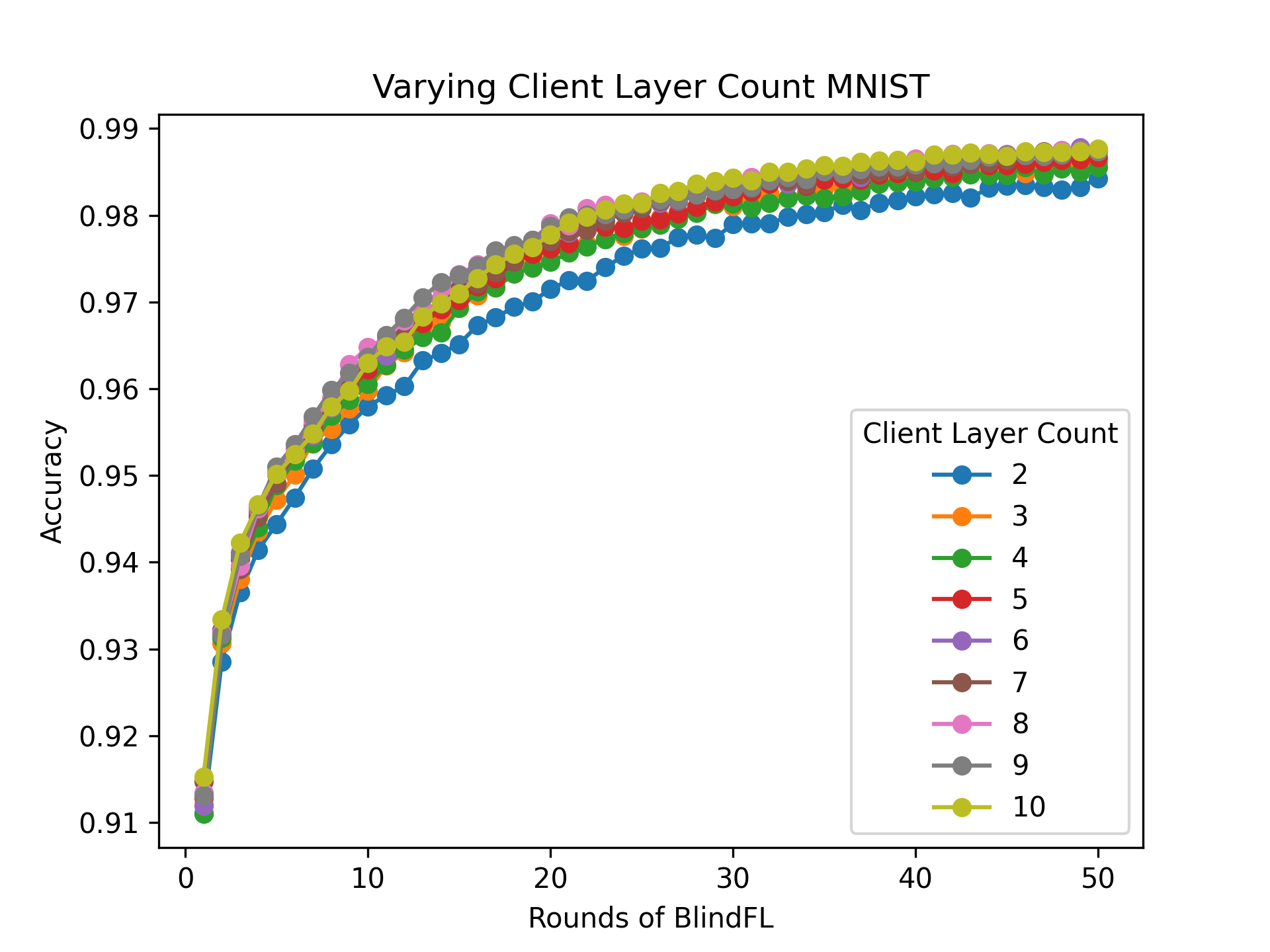}
  \caption{Accuracy averaged over 5 runs for varying client parameter matrix counts over multiple rounds for a LeNet-5 model trained on the MNIST dataset in a BlindFL federation.}
  \label{fig:mnist_clients_3}
  \vspace*{-\baselineskip}

\end{figure}

\begin{figure}[h]
  \centering
  \includegraphics[width=\linewidth]{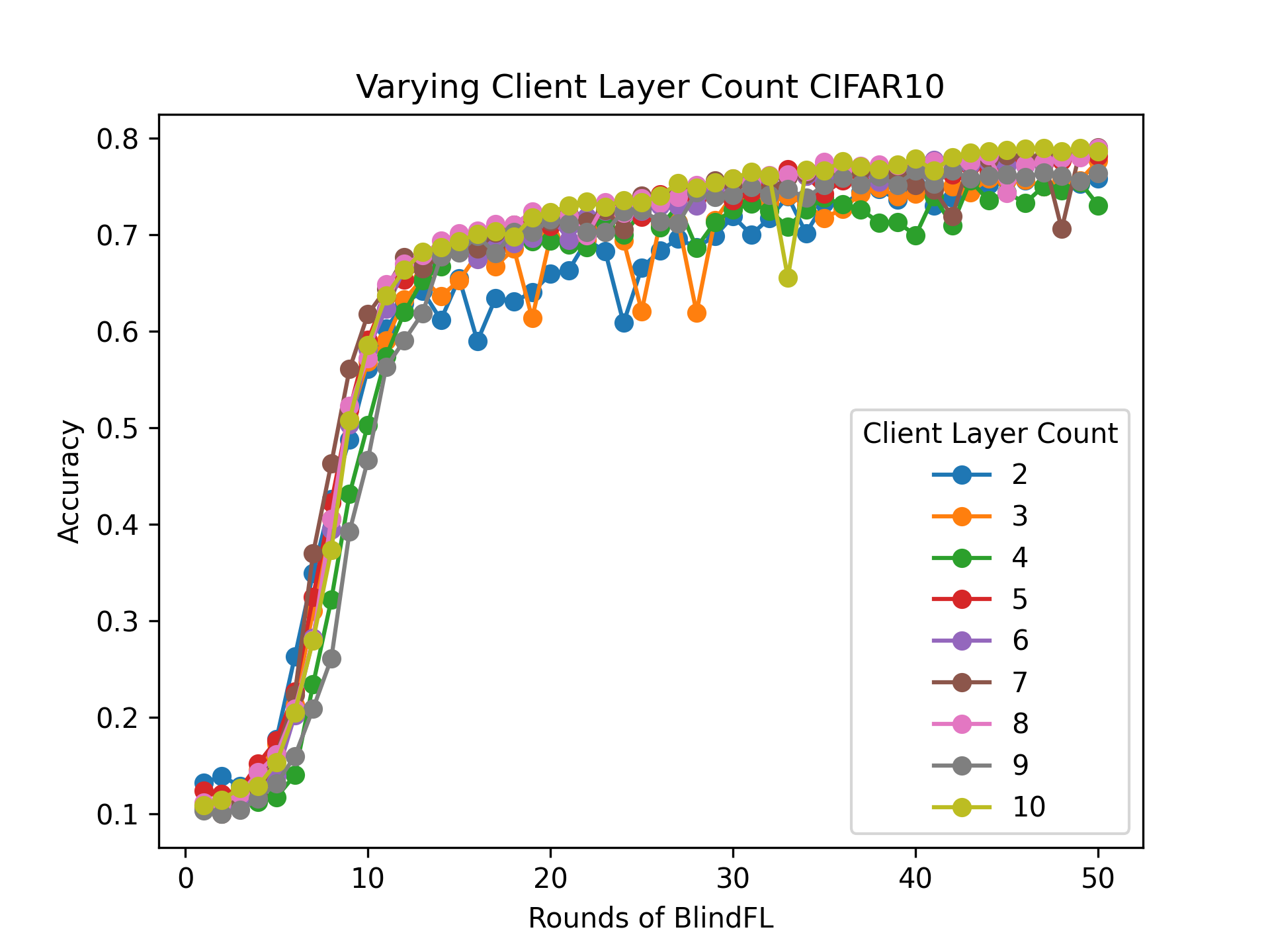}
  \caption{Accuracy averaged over 5 runs for varying client parameter matrix counts over multiple rounds for a ResNet-20 model trained on the CIFAR-10 dataset in a BlindFL federation.}
  \label{fig:cifar_clients_3}
  \vspace*{-\baselineskip}

\end{figure}

Figures \ref{fig:mnist_clients_3} and \ref{fig:cifar_clients_3} show, for both MNIST and CIFAR-10 datasets, how round-over-round performance is affected by changing the number of client parameter matrices collected per each global model parameter matrix for aggregation. All runs of the simulation for this experiment had 10 total clients and the FHE component was active.

Notice that even when only 2 parameter matrices are collected from each client, the performance of the global model is nearly the same as when all parameter matrices are collected from every client. For the specific, numeric impacts to LeNet-5 model test accuracy over MNIST and CIFAR, see Table \ref{table:3}. We experimentally conclude that CMS, even when $20$ percent of the model is sent from each client, has minimal adverse effect to performance.

\begin{table}[h!]
\begin{center}
\begin{tabular}{ |c|c|c| }
    \hline

    \textbf{\thead{Percent of Client Model \\ Param. Matricies Shared}} & \textbf{\thead{MNIST Test  \\ Accuracy}} & \textbf{\thead{CIFAR Test \\ Accuracy}} \\
    \hline\hline
 20\% Client Model & 98.43\% & 75.85\% \\ 
 30\% Client Model & 98.61\% & 77.71\% \\  
 40\% Client Model & 98.55\% & 73.06\% \\  
 50\% Client Model & 98.67\% & 78.20\% \\ 
 60\% Client Model & 98.75\% & 78.95\% \\  
 70\% Client Model & 98.76\% & 79.05\% \\   
 80\% Client Model & 98.76\% & 78.97\% \\ 
 90\% Client Model & 98.73\% & 76.42\% \\  
 100\% Client Model & 98.77\% & 78.68\% \\  
\hline
\end{tabular}
\end{center}
\caption{Final recorded accuracies for Figure 7 and 8.}
\label{table:3}
\vspace*{-\baselineskip}
\end{table}

\subsection{Effects of FHE and CMS}

\begin{figure}[h]
  \centering
  \includegraphics[width=\linewidth]{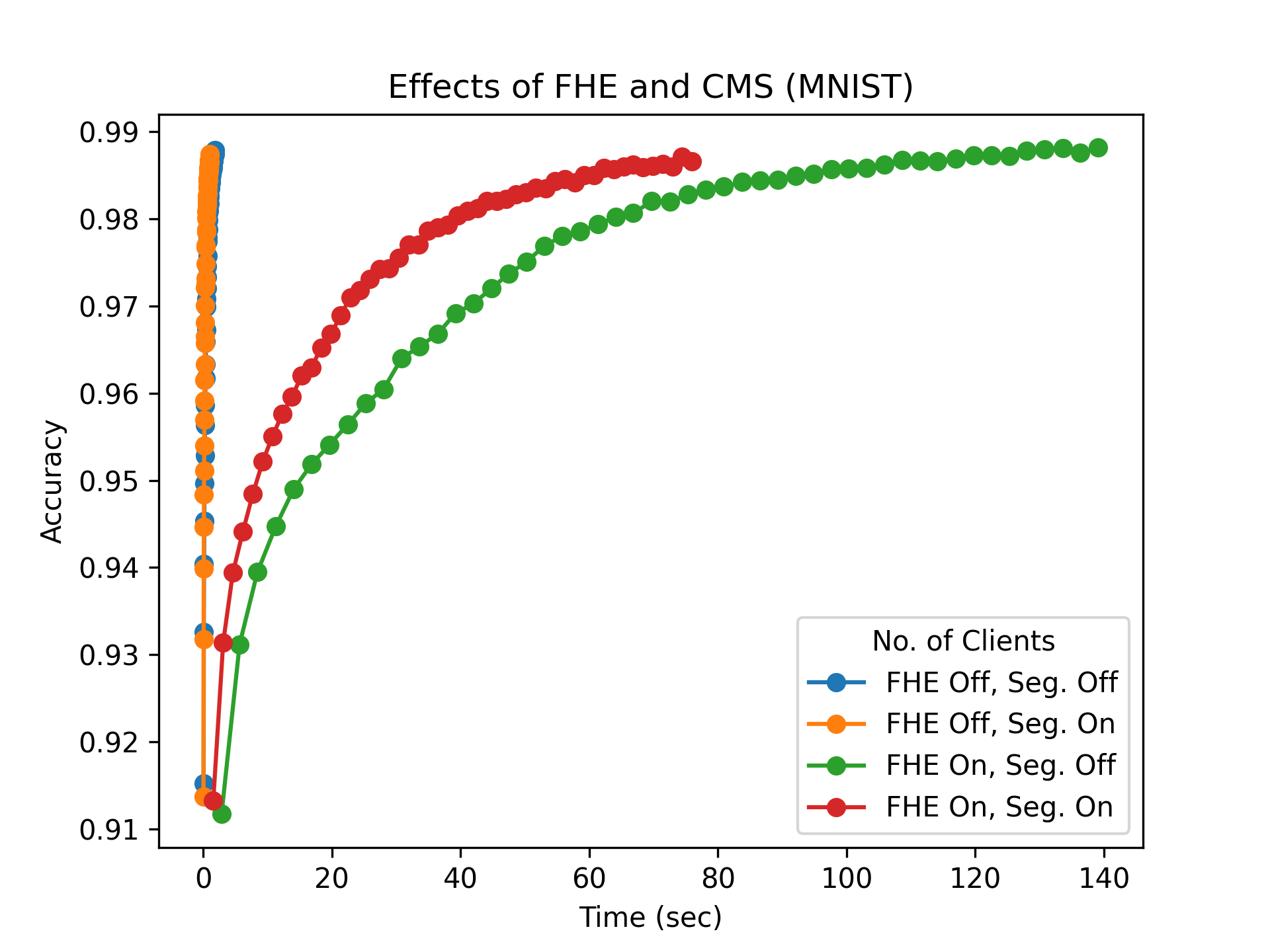}
  \caption{Accuracy averaged over 5 runs over multiple rounds for a LeNet-5 model trained on the MNIST dataset under different federation types where segmentation uses 5 parameter matrices per client.}
  \label{fig:mnist_client_4}
\end{figure}

\begin{figure}[h]
  \centering
  \includegraphics[width=\linewidth]{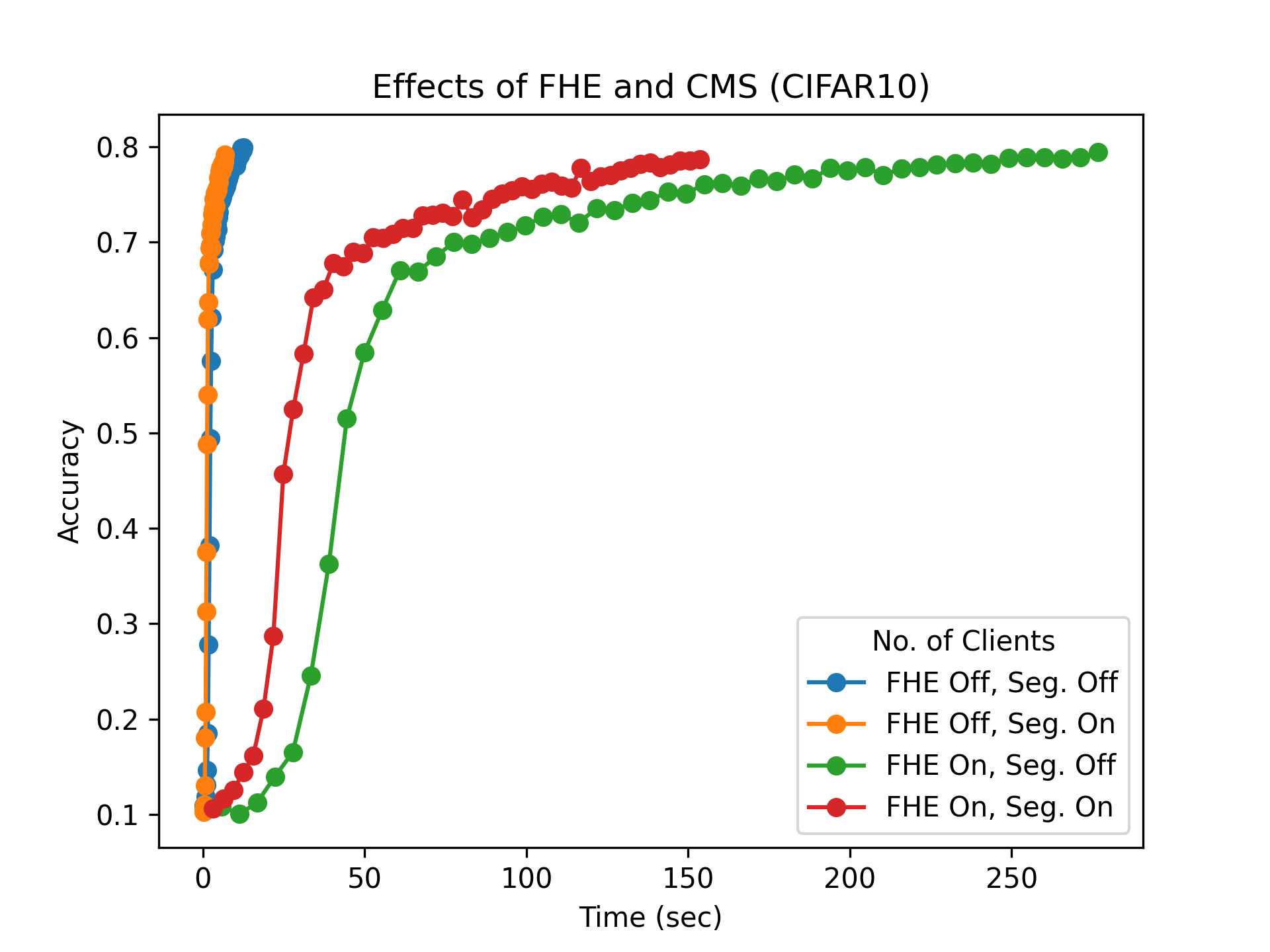}
  \caption{Accuracy averaged over 5 runs over multiple rounds for a ResNet-20 model trained on the CIFAR-10 dataset under different federation types where segmentation uses 50\% matrices available per client.}
  \label{fig:cifar_client_4}
\end{figure}

Figures \ref{fig:mnist_client_4} and \ref{fig:cifar_client_4} show, for both MNIST and CIFAR-10, the round-over-round performance and 5 run accuracy against server-side processing time for four different run types of the simulation:

\begin{itemize}
    \item \textbf{Standard FL}: An experiment run without FHE and without CMS.
    \item \textbf{FL+FHE}: An experiment run with FHE and without CMS.
    \item \textbf{FL+CMS}: An experiment run without FHE and with CMS.
    \item \textbf{BlindFL}: An experiment run with FHE and with CMS.
\end{itemize}

The number of clients for these runs was set to 10 where each data point represents one round of a given experiment. For Figures 11 and 12, time is measured as the amount of time spent by the server to aggregate client model updates. Encryption, decryption, and transfer time are not taken into consideration for these figures.

Over the course of 50 rounds of FL, we see that the same level of accuracy is achieved across all four run types, for both MNIST and CIFAR-10, highlighted again in Tables \ref{table:4}—\ref{table:cifar-7}. In other words, while FHE introduces a clear time overhead, neither component of BlindFL notably impacts the level of accuracy that is achieved. I.e., segmentation does not meaningfully decrease convergence accuracy but does reduce the amount of time needed to carry out FL.

\begin{table}[h!]
\begin{center}
\begin{tabular}{ |c|c|c| }
 \hline Experiment
  & MNIST Test Accuracy & Trial Time (Sec.)\\
 \hline\hline
 FHE Off, Seg. Off & 98.79\% & 1.80s\\ 
 FHE Off, Seg. On & 98.72\% & 1.00s\\  
 FHE On, Seg. Off & 98.82\% & 139.14s\\  
 FHE On, Seg. On & 98.66\% & 75.97s\\ 
\hline
\end{tabular}
\end{center}
\caption{For our different federations types, the 50th round timing and accuracy for Figure 9.}
\label{table:4}
\vspace*{-\baselineskip}
\end{table}

\begin{table}[h!]
\begin{center}
\begin{tabular}{ |c|c|c| }
 \hline Experiment
  & MNIST Test Accuracy & Trial Time (Sec.)\\
 \hline\hline
 FHE Off, Seg. Off & 98.71\% & 1.80s\\ 
 FHE Off, Seg. On & 98.56\% & 0.65s\\  
 FHE On, Seg. Off & 98.82\% & 139.18s\\  
 FHE On, Seg. On & 98.53\% & 49.94s\\ 
\hline
\end{tabular}
\end{center}
\caption{For our different federations types, the accuracy and timing averaged over 5 runs over 50 rounds for a LeNet-5 model trained on the MNIST dataset under BlindFL with 3 parameter matrices.}
\label{table:5}
\vspace*{-\baselineskip}
\end{table}

\begin{table}[h!]
\begin{center}
\begin{tabular}{ |c|c|c| }
 \hline Experiment
  & CIFAR-10 Test Accuracy & Trial Time (Sec.)\\
 \hline\hline
 FHE Off, Seg. Off & 79.50\% & 12.42s\\ 
 FHE Off, Seg. On & 77.20\% & 4.43s\\  
 FHE On, Seg. Off & 78.57\% & 275.91s\\  
 FHE On, Seg. On & 77.53\% & 102.31s\\ 
\hline
\end{tabular}
\end{center}
\caption{For our different federations types, the accuracy and timing averaged over 5 runs over 50 rounds for a ResNet-20 model trained on the CIFAR-10 dataset under BlindFL with 30\% of available parameter matrices.}
\label{table:cifar-6}
\vspace*{-\baselineskip}
\end{table}

\begin{table}[h!]
\begin{center}
\begin{tabular}{ |c|c|c| }
 \hline Experiment
  & CIFAR-10 Test Accuracy & Trial Time (Sec.)\\
 \hline\hline
 FHE Off, Seg. Off & 79.92\% & 12.43s\\ 
 FHE Off, Seg. On & 79.14\% & 6.76s\\  
 FHE On, Seg. Off & 79.42\% & 276.79s\\  
 FHE On, Seg. On & 78.69\% & 153.65s\\ 
\hline
\end{tabular}
\end{center}
\caption{For our different federations types, the 50th round timing and accuracy for Figure 10.}
\label{table:cifar-7}
\vspace*{-\baselineskip}
\end{table}

\subsection{Amount of Data Sent per Client}

Table \ref{table:6} shows, for MNIST, the average amount of data sent to the server per client, varying the number of client parameter matrices CMS sends and setting FHE both on and off. We recognize that parameter matrices of a given model are of differing sizes; however, each client has an equal and random chance of having any given layer selected. Therefore, the average for a given count of client parameter matrices is appropriate. For this experiment, 10 clients were used, and we average memory size across all 10 clients. We calculate the average total bytes required by a single client. We note that the raw LeNet-5 model includes 2,183KB of data, and so it should make sense that 1 plaintext layer shared of a LeNet-5 model includes, on average, $218$KB of data, i.e. $218$KB $\times$ $10$ clients gives a full model of 2,183KB of data.

\begin{table}[h]
    \centering
    \begin{tabular}{|c|c|c|}
        \hline
        \textbf{Client Param. Matrices} & \textbf{Without FHE} & \textbf{With FHE} \\
        \hline\hline
        1 Layer Shared & 218 KB & 5977 KB \\
        2 Layer Shared & 436 KB & 11,955 KB \\
        3 Layer Shared & 655 KB & 17,933 KB \\
        4 Layer Shared & 873 KB & 23,753 KB \\
        5 Layer Shared & 1091 KB & 29,888 KB \\
        6 Layer Shared & 1280 KB & 35,237 KB \\
        7 Layer Shared & 1310 KB & 36,417 KB \\
        8 Layer Shared & 1528 KB & 41,922 KB \\
        9 Layer Shared & 1727 KB & 47,821 KB \\
        10 Layer Shared & 2183 KB & 59,777 KB \\
        \hline
    \end{tabular}
    \caption{Amount of data sent per client (MNIST).}
    \label{table:6}
\end{table}

\begin{table}[h]
    \centering
    \begin{tabular}{|c|c|c|}
        \hline
        \textbf{\thead{Percent of Client Model \\ Param. Matricies Shared}} & \textbf{\thead{Without FHE}} & \textbf{\thead{With FHE}} \\
        \hline\hline
        10\% Client Model & 115 KB & 12,348 KB \\
        20\% Client Model & 231 KB & 24,854 KB \\
        30\% Client Model & 346 KB & 36,889 KB \\
        40\% Client Model & 445 KB & 48,529 KB \\
        50\% Client Model & 577 KB & 61,350 KB \\
        60\% Client Model & 661 KB & 69,687 KB \\
        70\% Client Model & 765 KB & 80,306 KB \\
        80\% Client Model & 846 KB & 88,329 KB \\
        90\% Client Model & 924 KB & 98,868 KB \\
        100\% Client Model & 1154 KB & 122,701 KB \\
        \hline
    \end{tabular}
    \caption{Amount of data sent per client (CIFAR-10).}
    \label{table:7}
\end{table}

Table \ref{table:7} shows, for CIFAR-10, the average amount of data set to the server per client, again varying the number of client parameter matrices CMS sends and setting FHE both on and off. We recognize that ResNet-20, with 128 parameter matrices, contains far more, albeit smaller, parameter matrices. We instead show the percent of the model shared and the associated impact of BlindFL on the amount of data sent per client. Our ResNet-20 model, with only 284,426 parameters, is smaller than our LeNet-5 model.
However, since each paramater matrix $j$ is individually encrypted, the space impact to the ResNet-20 model is greater.

\subsection{CMS Protection Against Inversion Attacks}

\begin{figure}[h]
  \centering
  \includegraphics[width=\linewidth]{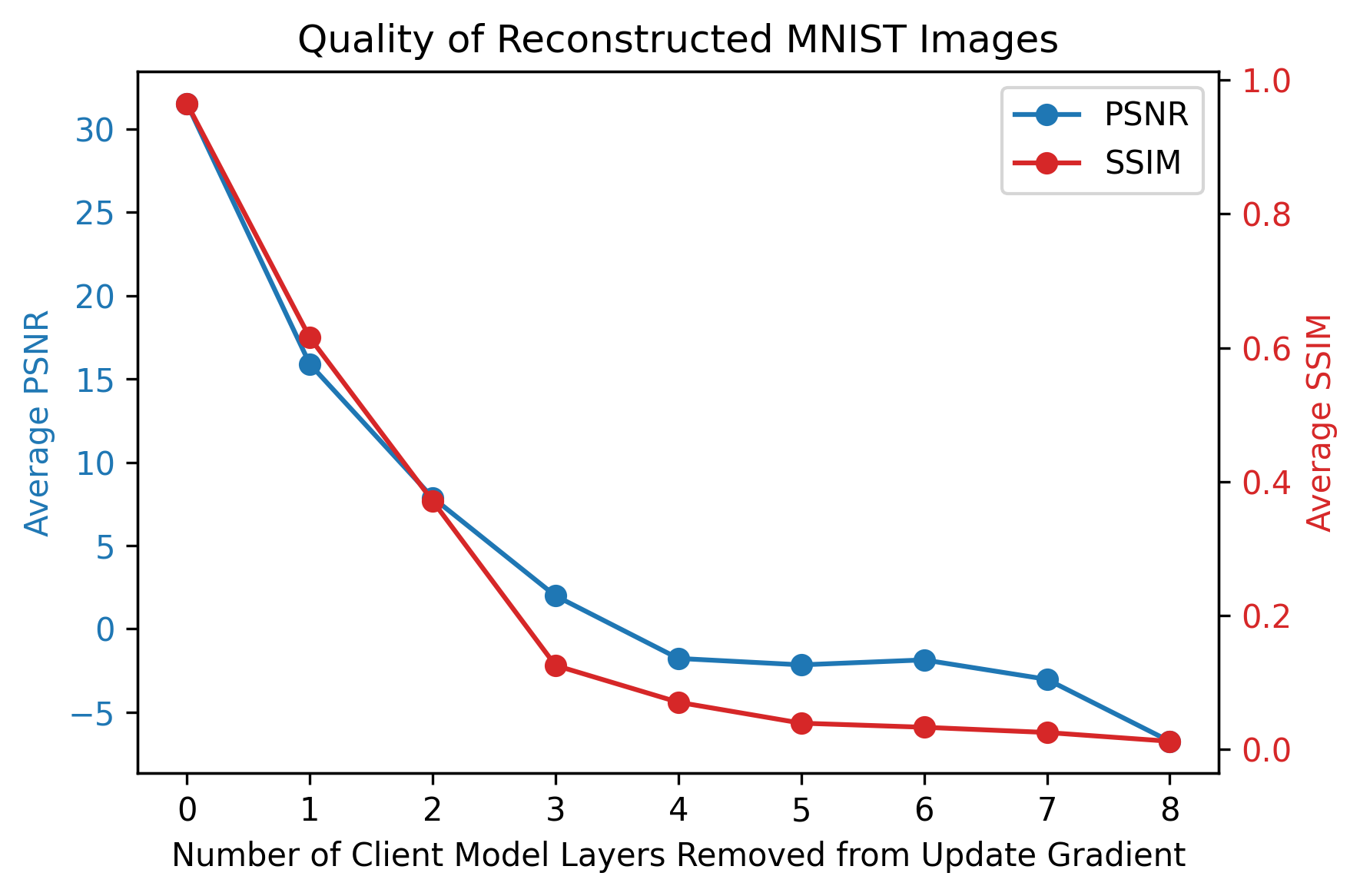}
  \caption{
  Average PSNR and SSIM over 15 GIA reconstructions varying the number of model layers represented in the attacked gradient for MNIST.
  }
  \label{fig:attack_1}
\end{figure}

Figures \ref{fig:attack_1} and \ref{fig:attack_2} show, for each of our datasets, how the number of client model layers present in an attacked model gradient affects the success of private data reconstruction by a GIA. For this experiment, we use the GIA described by Geiping, et al. \cite{inverting-gradient-attacks-easy}. We choose this GIA due to its minimal requirements on the adversary's knowledge of the underlying data distribution, only requiring estimates for the mean and standard deviation. Moreover, the attack is remarkably powerful even with these minimal requirements.

To assess the quality of the reconstructed images, we leverage both peak signal-to-noise ratio (PSNR) and structural similarity index measure (SSIM) \cite{wang2004image} as metrics. PSNR is a straight-forward function of the mean squared error between the ground truth and the reconstruction. SSIM, on the other hand, is a measure of the difference in structural information, taking into consideration luminance, contrast, and similarity between regions of pixels.

Recall adversary $\mathcal{A}$ from our threat model and the description of its model poisoning procedure from Section 5. For our experiments, suppose that $\mathcal{A}$ would like to recover private data from class $a$ of the main task. We consider the worst-case scenario where there is a single data sample $d_j$ labeled $a$ among all the training datasets of the uncompromised clients of the federation for round $i$. Let $d_j$ belong to the local dataset of client $c_j$, and let $\mathcal{A}$ send a malicious update for round $i$ which allows $\mathcal{A}$ to recover without error the gradient $g_j$ corresponding to $d_j$ from the decrypted global model for round $i+1$. I.e., for our experiments, let $\mathcal{A}$ perfectly recover the gradient contribution $g_j$ of $d_j$ from the global model gradient. We employ CMS for our experiments, and thus $g_j$ is a subset of the full gradient contribution $g'$ of $d_j$ to $c_j$'s local model $M_j$. After acquiring $g_j$, $\mathcal{A}$ then performs the GIA previously mentioned to reconstruct the target sample $d_j$ from $g_j$. Let the number of layers represented in the client model update sent by $c_j$ to the central server be $n$ and the total number of layers in $M_j$ be $N$. We investigate experimentally the effect of reducing $n$ on the ability of $\mathcal{A}$ to reconstruct $d_j$.

The methodology for Figure \ref{fig:attack_1} and \ref{fig:attack_2} only differ in the dataset used. For both figures, we generate the attacked gradient $g_j$ with an untrained ConvNet model whose design comes from Geiping, et al. \cite{inverting-gradient-attacks-easy}, a DNN with eight convolutional layers and one final linear layer. For our experiments, removing a ``layer'' from the attacked gradient means removing both the weight and the bias parameter matrix gradient for a given model layer from $g_j$. As for the GIA parameters, for each trial, i.e. reconstruction, we allow for three restarts and set max iterations to 4,000. For a trial in our experiments with $n$ layers represented, we randomly sample which $n$ layers to include in $g_j$ and use the above GIA to reconstruct an approximation of $d_j$. We average together the PSNR and SSIM values of reconstructions over $15$ trials per value of $n$, excluding when $n=0$.

\begin{figure}[h]
  \centering
  \includegraphics[width=\linewidth]{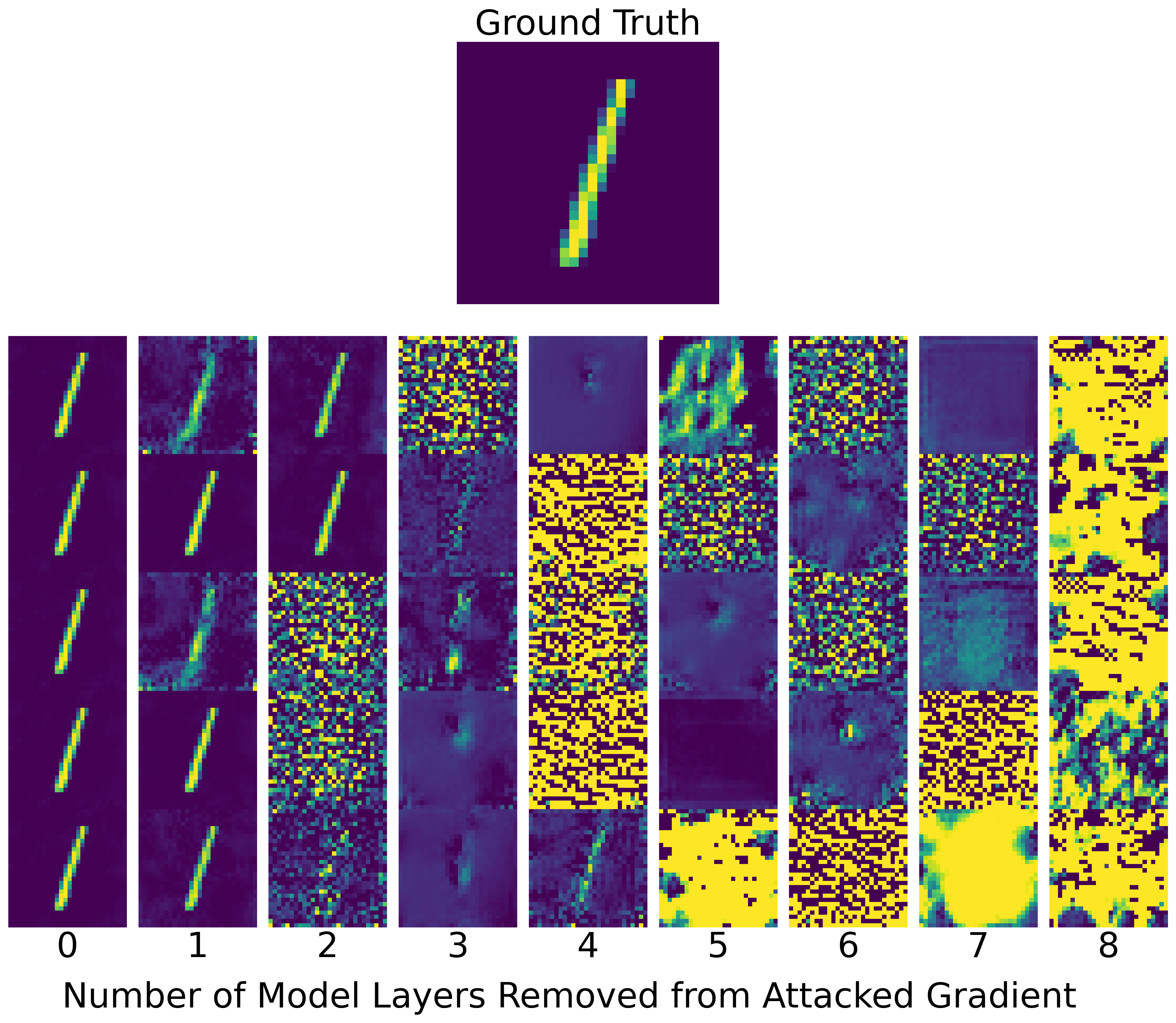}
  \caption{The first 5 of the 15 averaged GIA reconstructions from Figure \ref{fig:attack_1} as model layer gradients are removed from the attacked gradient.
  }
  \label{fig:mnist-attack}
\end{figure}

\begin{figure}[h]
  \centering
  \includegraphics[width=\linewidth]{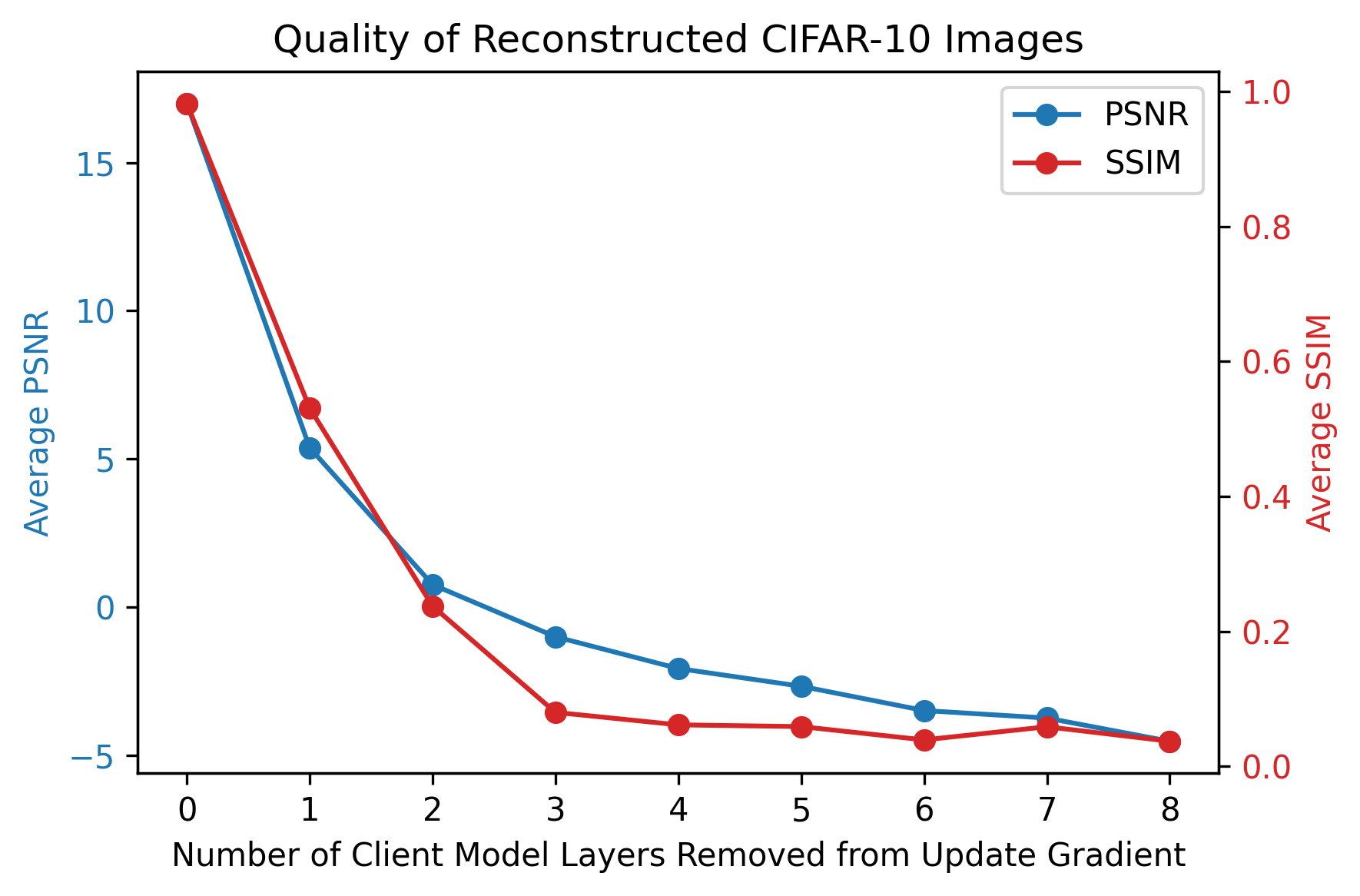}
  \caption{Average PSNR and SSIM over 15 GIA reconstructions varying the number of model layers represented in the attacked gradient for CIFAR-10.}
  \label{fig:attack_2}
\end{figure}

\begin{figure}[h]
  \centering
  \includegraphics[width=\linewidth]{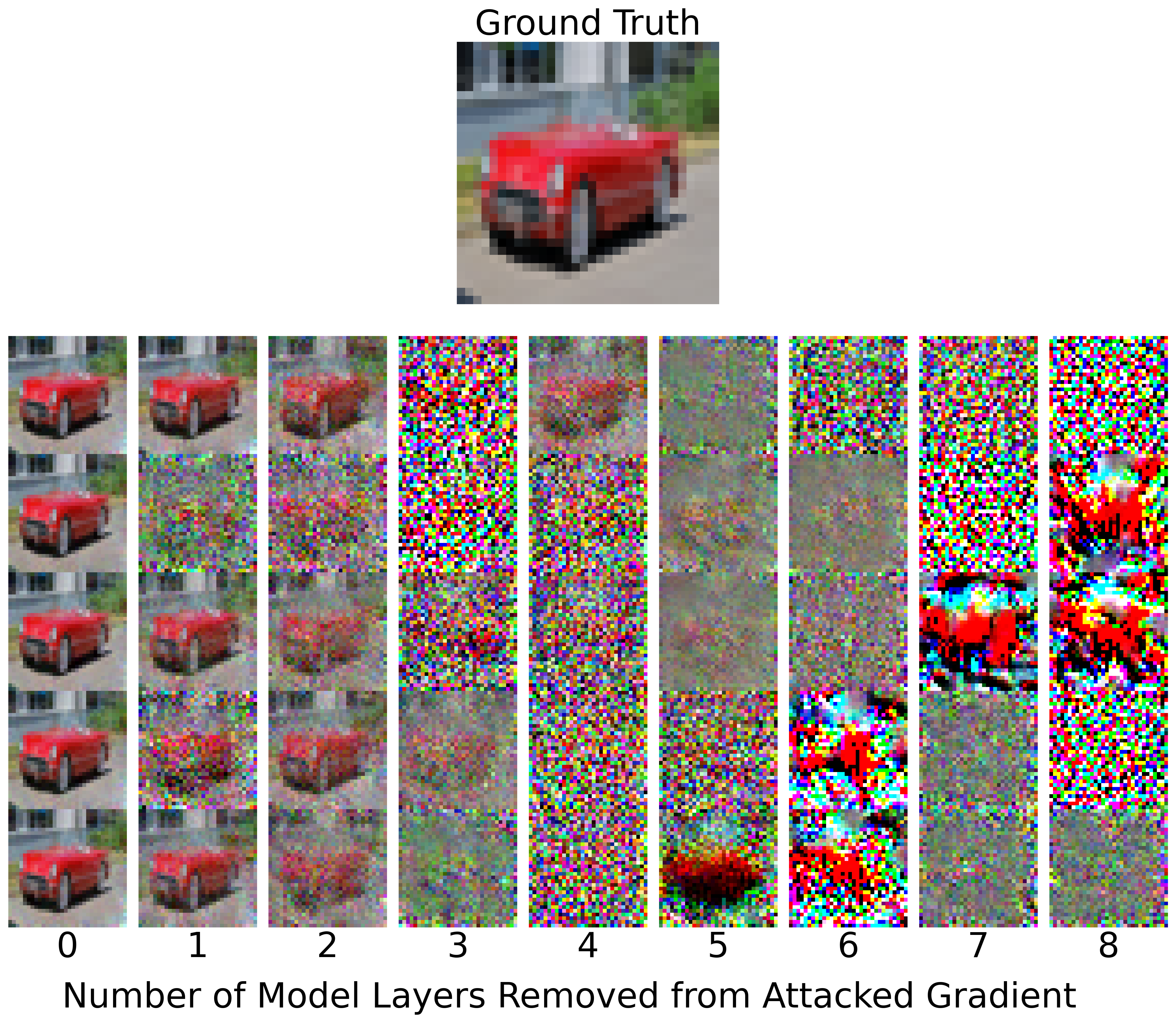}
  \caption{
  The first 5 of the 15 averaged GIA reconstructions from Figure \ref{fig:attack_2} as model layer gradients are removed from the attacked gradient.
  }
  \label{fig:cifar-attack}
\end{figure}

For both datasets, we see a stark drop in the performance of the GIA as $n$ decreases. Figures \ref{fig:mnist-attack} and \ref{fig:cifar-attack} display the reconstruction's impact visually over a decreasing number of layers represented in the gradient. Recall that we define $\mathcal{A}$ to have capabilities surpassing state-of-the-art malicious client attacks. Even in this worst-case scenario, we observe that the segmentation brought by CMS significantly hinders the success of an attack. Notice that for both Figures \ref{fig:attack_1} and \ref{fig:attack_2}, we see an exponentially decaying relationship across both metrics.

From the results of these experiments we get suggested values for our architecture hyperparameters. Let $r$ be the ratio of client layers included in the sent gradient, i.e. $r=\frac{n}{N}$. From our experiments, we see significant client-to-client protection when $r\leq2/3$. As for the number $p$ of parameter matrices to average per global model parameter matrix, to balance system and global model performance, we set $p=\frac{C}{2}$, where C is the total number of clients. Then, we get that the total number of clients that encrypt and send their model updates is $c\geq\left\lceil \frac{p}{r} \right\rceil$. For greater system performance, choose lower values for $p$.

\section{Conclusion}

FL is a technique for preserving data privacy for distributed ML-based systems. However, the threat to data privacy within the federation posed by GIAs is significant. While many different PPFL techniques have been proposed, each comes with its own inherent risks and flaws. A technology shown to be very effective at protecting data against these attacks is FHE. Encryption of client data before aggregation using FHE protects against server-side adversaries, but it does have several drawbacks: increased computation time, higher memory requirements, larger network packet sizes, and consequently, a longer system runtime. These drawbacks make using FHE alone less ideal for systems with high volume, low compute, or low bandwidth requirements, such as edge systems.

Our proposed solution, BlindFL, is a scalable PPFL that enhances an FHE approach with CMS. CMS significantly reduces the demands of FHE on both the clients and the server. We are thus able to implement BlindFL in contexts where an FHE-only approach would be too slow or too data-intensive for practical deployment. Additionally, while FHE thwarts server-side attacks, it does nothing against client-side attacks. We demonstrate that CMS provides significant security for this largely unaddressed attack type \cite{shi2024dealing, wei2023client}.

Future work, however, is needed to expand on this client-to-client security offered by CMS. One potential improvement to the CMS algorithm is to make the generation of the request matrix collaborative with clients, where the sensitivity of different layers of the client model is considered. A smart choice of which $n$ client parameter matrices are sent to the server might further steepen the exponentially decaying relationship between $n$ and the success of client-side GIA attacks. Additionally, work could be carried out to extend our BlindFL system to a decentralized FL setting, explore sending individual private FHE keys to clients, and investigate minimally small, sub-parameter-matrix CMS.

The addition of CMS can cut server-side processing time roughly in half without impacting model accuracy. While our experiments show an increase in client-server bandwidth requirements due to FHE, CMS effectively mitigates this overhead. Additionally, BlindFL maintains practically identical accuracy to non-FHE models on both the MNIST and CIFAR-10 datasets. We also demonstrate that CMS provides significant protection against malicious clients in the worst case with proper parameter choice. Thanks to all the enhancements it offers, BlindFL can be confidently offered as an approach towards enhanced FL privacy.

\bibliographystyle{ACM-Reference-Format}
\bibliography{references}

\end{document}